\documentstyle[12pt]{article}
\begin{document}
\font\bss=cmr12 scaled\magstep 0
\title{Brane World Models  And Darboux Transformations }
\author{ A.V. Yurov
\small\\ Theoretical Physics Department, \small\\ Kaliningrad
State University, Russia,  yurov@freemail.ru}
\date {}
\renewcommand{\abstractname}{\small Abstract}
\maketitle
\maketitle
\begin{abstract}
We consider a 5-D gravity plus a bulk scalar field, and with a
3-brane. The Darboux transformation is used to construct some
exact solutions. To do this we reduce the system of equations,
which describes the 5-D gravity and bulk scalar field to the
Schr\"odinger equation. The jump conditions at the branes lead to
the jump potential in the Schr\"odinger equation. Using the
Darboux transformation with these jump conditions, we offer a new
exact solution of the brane equations, which represents a
generalization of the Rundall-Sundrum  solution. For simplicity,
the main attention is focused on the case when Hubble root on the
visible brane is zero. However, the argument is given that our
method is valid in more realistic models with cosmological
expansion.
\end{abstract}
\thispagestyle{empty}
\medskip
\section{Introduction.}

Recently, E.E. Flanagan, S.-H.Tye and I. Wasserman [1] considered
the 5-D gravity plus a bulk scalar field with parallel 3-branes,
for which the action is
\begin{equation}
S=\int d^4xdy\sqrt{\mid
g\mid}\left(\frac{1}{2}R-\frac{(\nabla\phi^2)}{2}-V(\phi)\right)-
\sum_b\int_{y_b} d^4x\sqrt{\mid{\tilde g}^b\mid}\sigma_b(\phi),
\label{action}
\end{equation}
The system of units is chosen such that $\kappa=1,$ where $\kappa$
is 5-D gravitational coupling constant. The coordinates are
$(x^{\mu},y)$, for $0\le\mu\le 3$, the $b$th brane is located at
$y=y_b$, $g_{ab}$ is the 5-D metric and ${\tilde g}^b_{\mu\nu}$ is
the induced metric on the $b$th brane. The brane tensions
$\sigma_b$ and the potential $V$ are the functions of the bulk
scalar field $\phi=\phi(y)$. The authors of [1] seek solutions to
the field equations of the form
\begin{equation}
ds^2=dy^2+A(y)\left(-dt^2+e^{2Ht}\delta_{ik}dx^idx^k\right).
\label{interval}
\end{equation}
To solve the equations of motion, following the praxis in the
supergravity theory, they define a superpotential $W(\phi)$ by
${\dot\phi}\equiv W'(\phi)$, where the dot denotes the derivative
in $y$ and the prime -- in $\phi$, and $H$ is the effective Hubble
constant on the brane. If $H=0$  then one gets
\begin{equation}
V(\phi)=W'^2/8-W^2/6.
\label{superpotential}
\end{equation}

Using this equation, the authors of [1] have suggested a simple
way for obtaining solutions (see also [2, 3]). On the other hand,
equation (\ref{superpotential}) implies a supersymmetry connection
between $V$ and $W$, so one can write down the Darboux
transformation for these equations. This is because for
one-dimensional systems~\footnote{ $\phi=\phi(y),$ so this field
is the solution of an ODE.} the supersymmetry and Darboux
transformation are the same in a sense [4].  The aim of this work
is to use the Darboux transformation to obtain exact solutions of
the field equations, which can be derived by minimizing the action
(\ref{action}). For simplicity, the main attention is focused on
the models without the cosmological expansion, although our
approach also works in more realistic models with $H\ne 0$ (see
Sec.5).

One should be quite careful carrying the Darboux transformation
into the brane world. The reason is connected with the stick-slip
nature of fields: the jump conditions at the branes are
${\dot\phi}(y_b^+)-{\dot\phi}(y_b^-)=\sigma'_b(\phi_b)$. So one
must make sure that the Darboux transformation be compatible with
the jump condition.  However, this  can be done. Having done so,
we have constructed a new exact solution of the brane equations,
which is some sense generalizes the Rundall-Sundrum, henceforth RS
solution.

This paper is organized as follows. In section 2 we consider the
Darboux transformation and it's compatibility with the jump
conditions on the brane. In the next section we present the new
solution, which can be called the {\em dressed RS-brane}. The
$n=2$ Darboux iteration and the role of shape-invariant potentials
are discussed in section 4. In section 5 we show that the Darboux
transformation can work in the case with the cosmological
expansion and address some problems, characteristic of such
models. Some concluding remarks are given in section 6.

\section{Darboux transformation and jump condition}

If $H=0$  and $u\equiv {\dot A}/A$ then from (\ref{action}) and
(\ref{interval}) we get the following system ([1]),
\begin{equation}
{\dot
u}=-\frac{2{\dot\phi}^2}{3}-\frac{2}{3}\sum_b\sigma_b(\phi)\delta(y-y_b),\qquad
u^2=\frac{{\dot\phi}^2}{3}-\frac{2}{3}V(\phi).
\label{dvaurav}
\end{equation}
The third equation has the form
$$
\ddot\phi+2u\dot\phi=\frac{\partial V(\phi)}{\partial\phi}+\sum_b\frac{\partial\sigma_b(\phi)}{\partial\phi}\delta(y-y_b),
$$
and can be obtained from the system (\ref{dvaurav}), so it is
enough to consider one.  If we have a single brane, which is
located at $y=0$, then
$\sigma(\phi(y))\delta(y)=\sigma_0\delta(y),$ where
$\sigma_0=\sigma(\phi(0))$ (we assume that $\phi(0)\ne 0$).
Introducing $\psi=\psi(y)=A^2,$ we get the Schr\"odinger equation
\begin{equation}
\ddot\psi=\left(v(y)-\beta\delta(y)-\lambda\right)\psi,
\label{schrodinger}
\end{equation}
where
$$
v(y)=-\frac{8}{3}V(\phi(y))+\lambda,\qquad \beta=\frac{4\sigma_0}{3}.
$$
Here $\lambda$ should not be taken for the cosmological constant
(which is included into the potential $V(\phi)$), being the usual
spectral parameter, which we need in order to iterate the Darboux
transformation (see below).

What about of the inverse problem of restoration  $A(y)$,
$\phi(y)$ and $V(\phi)$, knowing $\psi(y)$, $v(y)$ and $\lambda$?
Let consider this problem for large values of $y,$ where the
$\delta$-function term in (\ref{schrodinger}) is omitted. First of
all, we assume that $v(y)$ is such that
\begin{equation}
\displaystyle{
v(y)={\rm const}+{\tilde v}(y),\qquad
\int_{-\infty}^{+\infty}dy\,\left(1+|y|\right)|{\tilde v}(y)<\infty.}
\label{potential}
\end{equation}
The condition (\ref{potential}) is generally unnecessary for the
use of the Darboux transformation, but in this paper we content
ourselves with such potentials only.

To reconstruct $A(y)$ from $\psi(y)$, one needs to ensure that
$\psi(y)>0$ for all $y$. The ground state wave function is the
only wave function in the physical spectrum, possessing this
property. Furthermore, it is the only solution of
(\ref{schrodinger}) (not only in the physical spectrum) which  can
be used to find the physical scalar field $\phi$ for the
potentials (\ref{potential}). The reason is as follows: if $\psi$
is the ground state wave function, then there exist such real
$p^2$ that
$$
\psi(y)\to e^{-p^2 y}\qquad {\rm{as}}\qquad y\to +\infty,
$$
so from the first equation (\ref{dvaurav}) one gets
$$
{\dot\phi}^2=-\frac{3 \dot u}{2}=-\frac{3}{4}\frac{d^2}{dy^2}\log\psi(y)\to \frac{3p^2}{4}>0\qquad
{\rm as}\qquad y\to +\infty.
$$
On the other hand, one can use $\psi$ as the solution of
(\ref{schrodinger}) with $\lambda<\lambda_{_{vac}}$. Such a
solution does not describe the bound state, but it can be positive
for any values of $y$ (one can choose the constants of integration
such that this function does not have zeros, so $A(y)$ can be
constructed starting out of this $\psi$), but in this case one
gets ${\dot\phi}^2<0$ as $y\to +\infty$.

{\em This is why throughout this paper we shall study only the
ground states of (\ref{schrodinger}).}

On the other hand, the Darboux transformation (henceforth DT)
allows one to construct potentials with any bound state {\em ad
hoc.} Therefore it is natural to consider the DT as a method for
constructing the exact solutions of (\ref{dvaurav}). This is the
crucial point of this paper.

Let $\psi_1=\psi(y;\lambda_1)$ and  $\psi_2=\psi(y;\lambda_2)$ are
two solutions of  (\ref{schrodinger}). We call $\psi_1$ the prop function for
the DT which has the form [5]:
\begin{equation}
\psi_2\to
\psi_2^{(1)}=\frac{{\dot\psi_2}\psi_1-{\dot\psi_1}\psi_2}{\psi_1},\qquad
v\to v^{(1)}=v-2\frac{d^2}{dy^2}\log\psi_1,\qquad
\lambda_2\to\lambda_2.
\label{dt}
\end{equation}
The DT (\ref{dt}) is an isospectral symmetry of
(\ref{schrodinger}): the dressed function $\psi_2^{(1)}$ is the
solution of the dressed equation (\ref{schrodinger}), namely
$$
\ddot\psi_2^{(1)}=\left(v^{(1)}-\beta^{(1)}\delta(y)-\lambda_2\right)\psi_2^{(1)},
$$
with the new (dressed) potential $v^{(1)}$ and the same eigenvalue
$\lambda^{(1)}_2=\lambda_2$.

The transformation law for the prop function is ([4])
\begin{equation}
\psi_1\to\psi_1^{(1)}=\frac{1}{\psi_1}\left(C_1+C_2\int^y \psi_1^2(y') dy'\right),\qquad
\ddot\psi_1^{(1)}=\left(v^{(1)}-\beta^{(1)}\delta(y)-\lambda_1\right)\psi_1^{(1)},
\label{selfdt}
\end{equation}
with arbitrary constants $C_{1,2}$ (and, of course the same
eigenvalue $\lambda^{(1)}_1=\lambda_1$).

What about the ground  state? If $\psi_1$ is the wave function of
the ground state for the initial potential, and $\lambda_2$ is the
next level in the physical spectrum, then after the DT (\ref{dt})
the dressed function $\psi^{(1)}_2$ will be the wave function of
the new ground state [6]. In this case, the spectrum of the
Hamiltonian with $v^{(1)}$ can be obtained by the deleting the
level $\lambda_1$ from the spectrum of  the initial Hamiltonian
(the case (i)). On the other hand, suppose
$\psi_1=\psi_1(y;\lambda_1)$ is not really the wave function of
the bound state of the initial Hamiltonian, but simply does not
have zeros and $\psi_1\to e^{+p^2 |y|}$ as $y\to\pm \infty$. Then
after (\ref{dt}) one gets a new Hamiltonian with the new ground
state with $\lambda=\lambda_1$ and the wave function
$\psi^{(1)}_1$ from the (\ref{selfdt}), with $C_2=0$ (the case
(ii)). The latter alternative is used throughout this paper.

The jump conditions at the brane are
\begin{equation}
\psi_{i+}(0)=\psi_{i-}(0)=\psi_i(0),\qquad {\dot\psi}_{i+}(0)-
{\dot\psi}_{i-}(0)=-\beta\psi_i(0).
\label{jump}
\end{equation}
where $\psi_{i\pm}(y)$ are the solutions of the Schr\"odinger
equation (\ref{schrodinger}) with the same potential but different
eigenvalues:
\begin{equation}
{\ddot\psi}_{i\pm}=\left(v(y)-\beta\delta(y)-\lambda_i\right)\psi_{i\pm}.
\label{schrodinger2}
\end{equation}

It is easy to see that conditions (\ref{jump}) are the same as the
jump condition from the [1]:
$$
u_+(0)-u_-(0)=\frac{1}{2}\left(\frac{{\dot\psi}_+(0)}{\psi_+(0)}-\frac{{\dot\psi}_-(0)}{\psi_-(0)}\right)=
\frac{{\dot\psi}_+(0)-{\dot\psi}_-(0)}{2\psi(0)}=-\frac{\beta}{2}=-\frac{2\sigma_0}{3}.
$$

The following theorem shows that the DT preserve the jump
conditions (\ref{jump})
\newline
\newline
{\bf Theorem.} Let
$\psi_i(y)=\left\{\psi_{i+}(y>0),\psi_{i-}(y<0)\right\}$ with
$i=1,2$ be the solutions of (\ref{schrodinger2}), such that the
jump conditions (\ref{jump}) are valid. If
$\psi_1=\left\{\psi_{_{1+}},\psi_{_{1-}}\right\}$ is the prop
function then the Darboux-dressed function
$\psi^{(1)}_2=\left\{\psi^{(1)}_{2+},\psi^{(1)}_{2-}\right\}$ (the
prop function for $\psi_{_{2\pm}}$ is $\psi_{_{1\pm}}$
respectively) save the same jump conditions (\ref{jump}) with new
``tension'' $\beta^{(1)}=-\beta$:
\begin{equation}
\psi^{(1)}_{2+}(0)=\psi^{(1)}_{2-}(0)\equiv\psi^{(1)}_2(0),\qquad {\dot\psi}^{(1)}_{2+}(0)-
{\dot\psi}^{(1)}_{2-}(0)=+\beta\psi_i(0).
\label{jump2}
\end{equation}
The proof is a straightforward calculation, which we leave to the
reader. As a hint, there are two useful formulas:
$$
{\dot\tau}_{1\pm}=v-\lambda_1-\tau_{1\pm}^2,\qquad
{\dot\psi}^{(1)}_{2\pm}=\left(\lambda_1-\lambda_2\right)\psi_{_{2\pm}}-\tau_{1\pm}\psi^{(1)}_{2\pm},
$$
where $\tau_{1\pm}={\dot\psi}_{1\pm}/\psi_{_{1\pm}}$.

Iterating this theorem, one can conclude that the n-fold DT
results in dressed functions, which retain the jump condition
(\ref{jump}) with $\beta^{(n)}=(-1)^n\beta$, provided that the $n$
prop solutions $\{\psi_{i\pm}\}$ ($i=1,..,n$) of the initial
Schr\"odinger equation (\ref{schrodinger2}) and one {\bf dressing}
solution $\psi_{n+1\pm}=\psi$ of the same equation, with the same
potentials but a different $\lambda$, satisfy the jump condition
(\ref{jump}). If this is the case, then one can use the Crum
formulas [7] to present the result as
\begin{equation}
A^{(n)}_{\pm}(y)=\sqrt{\frac{D^{(\pm)}_n(y)}{\Delta^{(\pm)}_n(y)}},\qquad
v^{(n)}_{\pm}(y)=v-2\frac{d^2}{dy^2}\log\Delta^{(\pm)}_n(y),
\label{crum}
\end{equation}
where
$$
\Delta^{(\pm)}_n(y)= \left|\begin{array}{cccc}
\psi^{_{[n-1]}}_{_{n\pm}}&\psi^{_{[n-1]}}_{_{n-1\pm}} &...&\psi^{_{[n-1]}}_{_{1\pm}}\\
\psi^{_{[n-2]}}_{_{n\pm}}&\psi^{_{[n-2]}}_{_{n-1\pm}} &...&\psi^{_{[n-2]}}_{_{1,\pm}}\\
\star\\
\star\\
{\dot\psi}_{_{n\pm}}&{\dot\psi}_{_{n-1\pm}} &...&{\dot\psi}_{_{1\pm}}\\
\psi_{_{n\pm}}&\psi_{_{n-1\pm}}&...&\psi_{_{1\pm}}
\end{array}
\right|,
$$
and  $D^{(\pm)}_n=\Delta^{(\pm)}_{n+1},$ with
$\psi_{n+1\pm}=\psi_{\pm}\to\psi^{(n)}_{\pm}=\left(A^{(n)}_{\pm}\right)^2,$
where $\psi^{[k]}\equiv d^k\psi/dy^k$. In particular, one can see
that the jump conditions from [1] are valid:
$$
u^{(n)}_+(0)-u^{(n)}_-(0)=\left(-1\right)^{n+1}\frac{2\sigma_0}{3},
$$
where $u^{(n)}_{\pm}={\dot A}^{(n)}_{\pm}/A^{(n)}_{\pm}$ and $\sigma_0$ is the tension on the initial brane.

We shall use the case (ii), for  which the ground state of the
potential $v^{(n)}_{\pm}$ is defined by the formula
\begin{equation}
\psi^{(n)}_{_{n\pm}}=\left(A^{(n)}_{\pm}\right)^2=\frac{\Delta^{(\pm)}_{n-1}}{\Delta^{(\pm)}_n}.
\label{vacuum}
\end{equation}

\section{Single dressed RS brane.}

To use the DT (\ref{dt}), (\ref{selfdt}) we start with a simple
initial potential $v=\mu^2=\;$const (i.e. with ${\tilde v}=0$, see
(\ref{potential})). This is the case of the Randall-Sundrum (RS)
model for which $V=-3\mu^2/8$. The solution $\psi_1$ of
(\ref{schrodinger}) with $\lambda_1=0$ is
\begin{equation}
\psi_{_{1\pm}}=a_{_{1\pm}} e^{\mu y}+b_{_{1\pm}} e^{-\mu y},\qquad
\psi_{_{1+}}=\psi_1(y>0), \qquad \psi_{_{1-}}=\psi_1(y<0),
\label{opor}
\end{equation}
with the real  positive constants $a_{_{1\pm}}>0$,
$b_{_{1\pm}}>0$.

From (\ref{jump}) one gets 
\begin{equation}
a_{_{1-}}=a_{_{1+}}+\frac{\beta(a_{_{1+}}+b_{_{1+}})}{2\mu},\qquad
b_{_{1-}}=b_{_{1+}}-\frac{\beta(a_{_{1+}}+b_{_{1+}})}{2\mu}.
\label{link}
\end{equation}
From  (\ref{link}) one can see that
$a_{_{1+}}+b_{_{1+}}=a_{_{1-}}+b_{_{1-}}$ and hence use the
following representation: 
\begin{equation}
a_{_{1\pm}}=r^2\sin^2\alpha_{_{\pm}},\qquad 
b_{_{1\pm}}=r^2\cos^2\alpha_{_{\pm}}.
\label{represent}
\end{equation}
Using (\ref{represent}) 
and (\ref{link}) one gets the compatibility condition as
\begin{equation}
-\sin^2\alpha_{_{+}}<\frac{\beta}{2\mu}<\cos^2\alpha_{_{+}},
\label{compactibility}
\end{equation}
so $|\beta|<2\mu$.

Therefore, to construct new exact solutions via the DT, one needs
to use (\ref{opor}) with the additional conditions (\ref{link})
and (\ref{compactibility}).

Now we choose (\ref{opor}) as the prop function. After the DT
(\ref{dt}) we get
$$
\mu^2-\beta\delta(y)\to \mu^2+v^{(1)}_{\pm}-\beta^{(1)}\delta(y),
$$
where
$$
\displaystyle{
v^{(1)}_{\pm}=-\frac{8\mu^2 a_{_{1\pm}}b_{_{1\pm}}}{\left(a_{_{1\pm}} e^{\mu y}+ b_{_{1\pm}} e^{-\mu
y}\right)^2},}
$$
with $v^{(1)}_+= v^{(1)}(y>0)$ and  $v^{(1)}_-= v^{(1)}(y<0)$. This potential has the jump on the
brane
$$
v^{(1)}_+(0)-v^{(1)}_-(0)=-4\mu\beta\left(
\frac{\beta}{2\mu}+\frac{a_{_{1+}}-b_{_{1+}}}{a_{_{1+}}+b_{_{1+}}}\right).
$$
which is zero if we choose the tension as
\begin{equation}
\beta=\frac{2\mu(b_{_{1+}}-a_{_{1+}})}{a_{_{1+}}+b_{_{1+}}}.
\label{nastr}
\end{equation}
If this is the case, then  $v_-^{(1)}$ can be obtained from
$v_+^{(1)}$ by transposing $a_{_{1+}}$ and $b_{_{1+}}$.

We choose  $\psi^{(1)}_{_{1\pm}}=1/\psi_{_{1\pm}},$\footnote{By
the way, we choose the constants of integration $C_1=1$ and
$C_2=0$ to be in the ground state.} therefore as $y\to 0$:
$$
{\dot\psi}^{(1)}_{_{1+}}(0)-
{\dot\psi}^{(1)}_{_{1-}}(0)=-\beta^{(1)}\psi^{(1)}_1(0)=+\beta\psi^{(1)}_1(0),
$$
as it should be, according to (\ref{jump2}).

Using $\psi^{(1)}_{_{1\pm}},$ one can calculate that the dressed
metric has the form
$$
A^{(1)}_{\pm}=\frac{1}{\sqrt{a_{_{1\pm}} e^{\mu y}+b_{_{1\pm}} e^{-\mu y}}},
$$
so $A^{(1)}(y)\to e^{-\mu y/2}/{\sqrt {a_{_{1+}}}},$ as
$y\to+\infty,$ and  $A^{(1)}(y)\to e^{\mu y/2}/{\sqrt
{b_{_{1-}}}},$ as $y\to-\infty$. Tweaking (\ref{nastr}), one gets
$a_{_{1+}}=b_{_{1-}}$ so
$$
A^{(1)}(y)\to \frac{1}{\sqrt{a_{_{1+}}}}e^{-\mu\mid y\mid/2}, \qquad
y\to \pm\infty.
$$
In this case, one can set $b_{_{1+}}=a_{_{1-}}=0$ to transform
$A^{(1)}$ into the exact RS solution of the system
(\ref{dvaurav}). We get the initial RS potential via $\beta\to
-\beta$.

Let us return to the case of general position.  Then
$$
\displaystyle{
\left({\dot\phi}^{(1)}_{\pm}\right)^2=\frac{3\mu^2 a_{_{1\pm}}b_{_{1\pm}}}{\left(a_{_{1\pm}} e^{\mu y}+ b_{_{1\pm}}
e^{-\mu y}\right)^2},}
$$
and a simple calculations yields the well-known sine-Gordon
potential $V^{(1)}(\phi_{\pm})$:
$$
V^{(1)}(\phi_{\pm})=-\frac{3\mu^2}{8}\cos\left[\frac{4}{\sqrt{3}}\left(\phi_{\pm}(y)-\phi_{0\pm}\right)\right],
$$
with
$$
\phi(y)_{\pm}=\phi_{0\pm}+\sqrt{3}\arctan\left(\sqrt{\frac{a_{_{1\pm}}}{b_{_{1\pm}}}}e^{\mu y}\right),
$$
where $\phi_{0\pm}$ are arbitrary constants.

At last one can calculate the quantity
$\left(\sigma_0^{(1)}\right)'$. The result is
$$
\left(\sigma_0^{(1)}\right)'={\dot\phi}_+^{(1)}(0)-{\dot\phi}_-^{(1)}(0)=
\mu\sqrt{3}\sin\left(\alpha_{_+}-\alpha_-\right)\cos\left(\alpha_{_+}+\alpha_-\right).
$$
Note that in the case of tweaking (\ref{nastr})
$$
\beta=2\mu\cos 2\alpha_+,\qquad \alpha_-=\alpha_++\frac{\pi}{2}\left(2N+1\right),
$$
where $N$ is integer.

\section{Two applications}

There are many possible  applications of these results. In this
section we briefly consider the $n=2$ dressing of the RS brane and
the role of shape-invariant potentials.
\newline
\newline
{\bf 1. Double dressed RS model}
\newline
Another solution of the same Schr\"odinger equation with $\lambda_2=\mu^2-\nu^2$ has the form
\begin{equation}
\psi_{_{2\pm}}=a_{_{2\pm}} e^{\nu y}+b_{_{2\pm}} e^{-\nu y},\qquad
\psi_{_{2+}}=\psi_1(y>0), \qquad \psi_{_{2-}}=\psi_1(y<0),
\label{dressing}
\end{equation}
Using the theorem from section 2, one assumes  that
\begin{equation}
a_{_{2-}}=a_{_{2+}}+\frac{\beta(a_{_{2+}}+b_{_{2+}})}{2\nu},\qquad
b_{_{2-}}=b_{_{2+}}-\frac{\beta(a_{_{2+}}+b_{_{2+}})}{2\nu}.
\label{link2}
\end{equation}
Using (\ref{dt}) for (\ref{dressing}), one gets 
\begin{equation}
\begin{array}{cc}
\displaystyle{
\psi^{(1)}_{_{2\pm}}=\frac{\Omega_{\pm}}{a_{_{1\pm}} e^{\mu y}+b_{_{1\pm}} e^{-\mu y}},}\\
\\
\Omega_{\pm}=(\nu-\mu)\left(a_{_{1\pm}}a_{_{2\pm}}e^{(\mu+\nu)y}-b_{_{1\pm}}b_{_{2\pm}}e^{-(\mu+\nu)y}\right)+
(\nu+\mu)\left(a_{_{2\pm}}b_{_{1\pm}}e^{(\nu-\mu)y}-a_{_{1\pm}}b_{_{2\pm}}e^{(\mu-\nu)y}\right).
\end{array}
\label{dressed}
\end{equation}
The function (\ref{dressed}) will be the prop function for the
second DT. In order to create a new ground  state with the
``energy`` $\lambda_2<0$, one should choose $\nu>\mu>0$,
$a_{_{2\pm}}>0$ but $b_{_{2\pm}}<0$. In this case
$\psi^{(1)}_{_{2\pm}}$ does not have any zeros (nor has it any
singularities). Taking into account (\ref{link2}), we conclude
that positive $a_{_{2\pm}}$ must be such that:
$$
\frac{a_{_{2+}}}{a_{_{2-}}}>\left(\frac{2\nu}{2\nu+\beta}\right)_{{\rm if}\,\,\beta>0},\qquad
\frac{a_{_{2-}}}{a_{_{2+}}}>\left(\frac{2\nu}{2\nu-\beta}\right)_{{\rm if}\,\,\beta<0}.
$$
If this is the case, then we can use (\ref{crum}) and
(\ref{vacuum}) to proceed with the second DT, yielding
$$ v^{(2)}_{\pm}=\mu^2-2\frac{d^2}{dy^2}\log\Omega_{\pm},\qquad
A^{(2)}_{\pm}=\sqrt{\frac{a_{_{1\pm}} e^{\mu y}+b_{_{1\pm}}
e^{-\mu y}}{\Omega_{\pm}}}.
$$
One can verify that we have arrived in a brane with the right jump
condition and tension $\sigma^{(2)}_0=+\sigma_0$. Note that the
potential $v^{(1)}$ has two levels with $\lambda_1=0$ and
$\lambda_2=\mu^2-\nu^2$.
\newline
\newline
{\bf 2. Shape-invariant potentials}

Another way to obtain exact soluble potentials via DT is to use
the shape invariants [8].  If an initial potential is a function
of $y$ and some free parameters $a_i$ : $v=v(y;a_i)$, and after
the DT one gets $v^{(1)}=v^{(1)}(y;a^{(1)}_i)$ then  $v$ is called
a shape-invariant (SP) potential.

SP-potentials are common in quantum mechanics. An example is the
harmonic oscillator [4]. We stress that exact soluble potentials
from [1] (for the models without cosmological expansion) are
SP-potentials too. We confine ourselves with considering three
examples from the above article, retaining the terminology.
\newline
{\em  A. Even superpotential.} In the case of DFGK model [9] we
get  (with $A(0)=1$):
$$
\log A(y)=-\frac{ay}{3}+g\left(1-e^{2by}\right).
$$
In this case
$$
v(y)=\frac{8}{3}gb(2a-3b)e^{2by}+16g^2b^2e^{4by}+\frac{4a^2}{9}.
$$
After the DT
$$
v\to v^{(1)}=v-4\frac{d^2}{dy^2}\log A,
$$
we get
$$
v^{(1)}(y)=\frac{8}{3}gb(2a+3b)e^{2by}+16g^2b^2e^{4by}+\frac{4a^2}{9}.
$$
Thus $v^{(1)}$ can be obtained from $v$ by the substitution $a\to
-a$ and $g\to -g$. It mean that $v(y)$ is an SP-potential.
\newline
{\em B. Odd superpotential.}
In this case we have
$$
\log A(y)=-ay-by^2.
$$
A calculation yields
$$
v(y)=4\left(2by+a\right)^2-4b.
$$
It is nothing but the harmonic oscillator and, of course an
SP-potential: $ v^{(1)}=v+const$.
\newline
{\em C. Exponential potential.}
Superpotential is $W(\phi)=2b e^{-\phi}$, where
$\phi(y)=\log(a-by)$. Therefore
$$
v(y)=\frac{c}{(by-a)^2}.
$$
It is a well known SP potential $v^{(1)}=const\times v$.

\section{On models with cosmological expansion}

If $H\ne 0$ in (\ref{interval}) then we have more complex system instead of (\ref{dvaurav}) (in the case of the single
brane at $y=0$):
$$
{\dot
u}=-\frac{2{\dot\phi}^2}{3}-\frac{2H^2}{A}-\frac{2}{3}\sigma_0\delta(y),\qquad
u^2=\frac{{\dot\phi}^2}{3}-\frac{2}{3}V(\phi)+\frac{4H^2}{A}.
$$
Introducing $\psi=A^2,$ one can still reduce this system to the
Schr\"odinger equation
$$
{\ddot\psi}=\left(U(y)-\beta\delta(y)-\lambda\right)\psi,
\label{nschrodinger}
$$
where
\begin{equation}
\displaystyle{
U(y)=-\frac{8}{3}V+\frac{12H^2}{\sqrt{\psi}}+\lambda.}
\label{U}
\end{equation}
The reader may think that it is a very naive step, however this is
not the case! In fact, for any given $U(y)$ one can solve
(\ref{nschrodinger}) to find $\psi$ and find $V$ from (\ref{U}).
It is clear that solving (\ref{nschrodinger}) we should obtain a
positive $\psi$ in order to find $V$, but it is not a problem, as
has been demonstrated in sections 3 and 4. Indeed, one can
construct a positive solution via the DT starting out from any
simple solvable model, such as the RS model for example.

But here we face with another problem. To elucidate it, let us
start out with an ``RS potential'' $U=\mu^2$, and consider the case
$\lambda=0$. It is clear that the solution has the form
(\ref{opor}). Using it as the prop function in (\ref{dt}), one
gets the same metric $A^{(1)}_{\pm}$ from section 3 and, of
course, with another potential. The main problem now is the
kinetic energy of the scalar field, having assumed the form
$$
\displaystyle{
\frac{({\dot\phi^{(1)}_{\pm}})^2}{2}=\frac{3\mu^2a_{_{\pm}}b_{_{\pm}}}{2\left(a_{_{\pm}}e^{\mu y}+b_{_{\pm}}e^{-\mu y}\right)^2}-
\frac{3H^2}{2}\sqrt{a_{_{\pm}}e^{\mu y}+b_{_{\pm}}e^{-\mu y}}.}
$$
It is clear that for large enough $y$, one gets $\dot\phi^2<0$. On
the other hand, if (see (\ref{represent})
$$
\left(\frac{H}{\mu}\right)^2<\frac{\sin^22\alpha_{\pm}}{4r},
$$
then on the brane and near it, there is no trouble. Therefore a
simple way to avoid this  problem is to deal with the finite
volume, which is less then the value of $y$ resulting in the
negative kinetic term. Then one can use the DT and produce exact
solutions in the brane world with cosmological expansion as well.

\section{Conclusion}

This approach can be easy generalized to the case of $K$-branes.
To do it, we need to replace the term $\beta\delta(y)$ in
(\ref{schrodinger}) by
$$
\sum_{b}\beta_b\delta(y-y_b),
$$
and replace the two jump conditions (\ref{jump}) by $2K$ jump
conditions at $y=y_b$. This is technical matter, and the detail
has been omitted.

Another kettle of fish is that in the case of general position one
cannot reconstruct the form of the potential $V(\phi)$, as a
function of $\phi$. Nevertheless, all these solutions will have
the RS asymptotics, if constructed as has been described in
section 3. The reader can easily obtain a great many exact
solutions with the RS asymptotics, using the formulae from this
section.~\footnote{ It is interesting that in order to calculate
the n-times dressed Ricci scalar, one can use the formula
$$
R^{(n)}_{\pm}=-\frac{4 {\ddot A}^{(n)}_{\pm} A^{(n)}_{\pm}+ ({\dot
A}^{(n)}_{\pm})^2}{( A^{(n)}_{\pm})^2}+ \frac{12H^2}{A^{(n)}_{\pm}}.
$$
}
The DT gives a link between solvable problems, and one finds that
most of (if not all) the exactly solvable potentials, from the
harmonic oscillator to the finite-gap potentials [10], can be
obtained via these transformations. The physical sense of these
potentials in the brane world is not clear. Our main purpose here
has been merely to demonstrate and advertise the DT as a powerful
tool to manufacture exactly soluble potentials in the 5-D gravity
with the bulk scalar field. It has been demonstrated that this
method can be useful both in models without and with the
cosmological expansion, although in the latter case the situation
is more complex and not quite clear.

There is notably one imperfection of this method: the DT would
work for  the 5-D gravity only. If $D>5$ then the warp factor is a
multi-variable function. Unfortunately, there is no cogent and
effective general DT theory in several dimensions [11].

But if  $D=5$ (and maybe $H=0$) the DT  is probably the best way
to study the exactly solvable potentials.
$$
{}
$$
{\em Acknowledgements.} The author are grateful to D. Vassilevich
for useful comments, to M. Rudnev for his help, the Helmholtz program for financial support
and M. Bordag for his kind hospitality at the University of
Leipzig. I'd like to thank an anonymous referee, who has pointed
out a mistake in the earlier version of this paper.
$$
{}
$$
\centerline{\bf References} \noindent
\begin{enumerate}
\item E.E. Flanagan, S.-H.H. Tye and I. Wasserman,
\rm\, [hep-th/0110070 v2].
\item L. Randall and R. Sundrum, \rm\, Phys. Rev. Lett. {\bf 83},
3370 (1999), [hep-th/9906064].
\item S.-H.H. Tye and I. Wasserman,\rm\, Phys. Rev. let. {bf 86}
1682 (2001), [hep-th/0006068].
\item  V.B. Matveev and M.A. Salle.\rm\,  Darboux Transformation and
    Solitons. Berlin--Heidelberg: Springer Verlag, 1991.
\item  J.G. Darboux \rm\, Compt.Rend., 94 p.1343
(1882).
\item A.A. Andrianov, N.V. Borisov and M.V. Ioffe,\rm\,Phys.Lett.{\bf A(105)}
19 (1984); A.A. Andrianov, N.V. Borisov M.I. Eides and M.V. Ioffe,\rm\,Phys.Lett.{\bf A(109)}
143 (1984).
\item M.M. Crum \rm\, Quart. J. Math. Ser. Oxford 2, 6 p. 121
(1955).
\item L. Infeld and T.E. Hull, \rm \, Rev.Mod.Phys.  V. 23 p.
21 (1951).
\item O. DeWolf, D.Z. Freedman, S.S. Gubser and A. Karch,\rm\,
Phys. Rev. {\bf D62}, 046008 (2000), hep-th/9909134.
\item A.P. Veselov and A.B. Shabat,\rm\, Funkts. Anal. Prilozhen.,
Vpl. 27, No. 2, p. 1 (1993).
\item A. Gorizales-Lopes and N. Kamran, J.Geom.Phys.{\bf 26}202-226 (1998)
[hep-th/9612100].

\smallskip

\end{enumerate}

\end{document}